# Tightly-bound and room-temperature-stable excitons in van der Waals degenerate-semiconductor $Bi_4O_4SeCl_2$ with high charge-carrier density

Yueshan Xu[1,2,#], Junjie Wang[1,2,#], Bo Su[1], Jun Deng[1], Cao Peng[3], Chunlong Wu[3], Qinghua Zhang[1], Lin Gu[4], Jianlin Luo[1,5], Nan Xu[3,6,*], Jian-gang Guo[1,5,*], and Zhi-Guo Chen[1,2,5,*]

[1]Beijing National Laboratory for Condensed Matter Physics and Institute of Physics, Chinese Academy of Sciences, Beijing 100190, China

[2]School of Physical Sciences, University of Chinese Academy of Sciences, Beijing 100190, China

[3]Institute for Advanced Studies, Wuhan University, Wuhan, Hubei 430072, China

[4]State Key of Laboratory of New Ceramics and Fine Processing, School of Materials Science and Engineering, Tsinghua University, Beijing 100084, China

[5]Songshan Lake Materials Laboratory, Dongguan, Guangdong 523808, China

[6]Wuhan Institute of Quantum Technology, Wuhan, Hubei 430206, China

[#]Y.X. and J.W. contributed equally to this work.

*Email: zgchen@iphy.ac.cn (Z.G.C.), jgguo@iphy.ac.cn (J.G.G.), nxu@whu.edu.cn (N.X.).

## Abstract

**Excitons, which represent a type of quasi-particles consisting of electron-hole pairs bound by the mutual Coulomb interaction, were often observed in lowly-doped semiconductors or insulators. However, realizing excitons in the semiconductors or insulators with high charge carrier densities is a challenging task. Here, we perform infrared spectroscopy, electrical transport, ab initio calculation, and angle-resolved-photoemission spectroscopy studies of a van der Waals degenerate-semiconductor $Bi_4O_4SeCl_2$. A peak-like feature (i.e., $\alpha$ peak) is present around ~ 125 meV in the optical conductivity spectra at low temperature $T$ = 8 K and room temperature. After being excluded from the optical excitations of free carriers, interband transitions, localized states and polarons, the $\alpha$ peak is assigned as the exciton absorption. Moreover, assuming the existence of weakly-bound excitons—Wannier-type excitons in this material violates the Lyddane-Sachs-Teller relation. Besides, the exciton binding energy of ~ 375 meV, which is about an order of magnitude larger than those of conventional semiconductors, and the charge-carrier concentration of ~ $1.25 \times 10^{19}$ $cm^{-3}$, which is higher than the Mott density, further indicate that the excitons in this highly-doped system should be tightly bound. Our results pave the way for developing the optoelectronic devices based on the tightly-bound and room-temperature-stable excitons in highly-doped van der Waals degenerate semiconductors.**

## Introduction

Excitons have attracted intensive attention because they not only can provide a footstone for discovering novel quantum phenomena[1-4], e.g., exciton condensation[5-7], but also can play a crucial role in developing new-generation optoelectronic devices, such as excitonic light-emitting diodes[8-10]. It is well known that injecting free charge carriers into semiconductors or insulators can profoundly influence the dielectric background and partially fill the unoccupied single-particle states, which results in the screening of the exciton binding energy and the weakening of the excitonic feature intensity[11]. As the charge-carrier concentration becomes higher than the so-called Mott density ($n_M$), the occurrence of an insulator-to-metal transition is expected to be accompanied with the complete suppression of the excitonic feature[12,13]. Thus, it is challenging to realize excitons in highly-doped semiconductors or insulators[14-16]. Tightly-bound excitons host large binding energies which are about more than one order of magnitude larger than those in the majority of conventional quasi-2D semiconductor systems including group-III-V-based semiconductor quantum wells and thus have a larger probability to survive in highly-doped semiconductors or insulators[17]. Besides, the room-temperature stability of excitons is one of the key factors for their wide and realistic applications in optoelectronic devices[18-20]. It is worth noticing that van der Waals materials, which are usually characterized with strong in-plane bonds and weak out-of-plane van der Waals interactions, can provide a fertile ground for exploring exceptional phenomena because unexpected effects may be achieved by a variety of tunable methods, e.g., reducing the material thickness[21-23]. Moreover, the relatively high carrier concentration in van der Waals materials may enhance the photoresponsivity of optoelectronic devices[24]. Although (i) the large binding energies may make tightly-bound excitons have relatively large survival probabilities in highly-doped semiconductors or insulators, (ii) the room-temperature stability of excitons is significant for the wide and realistic applications, (iii) van der Waals materials are good contenders for realizing novel effects via tunable methods, and (iv) the high carrier doping in van der Waals materials may make the photoresponsivity of optoelectronic devices higher, tightly-bound and room-temperature-stable excitons were seldom found experimentally in highly doped van der Waals semiconductors or insulators. Therefore, searching for tightly-bound excitons stable at room temperature in highly doped van der Waals semiconductors or insulators is one of the frontier areas in condensed matter science.

Recently, a new van der Waals material, $Bi_4O_4SeCl_2$, was successfully synthesized[25]. In the unit cell of $Bi_4O_4SeCl_2$, the two antifluorite $[Bi_2O_2]^{2+}$ layers are connected by a $Se^{2-}$ layer and are covered with two $[Cl]^{-}$ layers, which can be regarded as a 1:1 intergrowth between $Bi_2O_2Cl_2$ and $Bi_2O_2Se$ slabs. Thermal transport experiments indicate an ultralow room-temperature thermal conductivity in $Bi_4O_4SeCl_2$, which suggests that $Bi_4O_4SeCl_2$ has a great potential for thermoelectric applications[26,27]. Electrical transport measurements show a high charge-carrier mobility in $Bi_4O_4SeCl_2$[25]. Furthermore, electrical transport study revealed that the temperature dependence of the resistivity of $Bi_4O_4SeCl_2$ mainly arises from the temperature evolution of the charge-carrier mobility in $Bi_4O_4SeCl_2$, rather than its charge-carrier concentration, which identifies $Bi_4O_4SeCl_2$ as a degenerate semiconductor[25]. Importantly, it was theoretically predicted that the bound excitons with large binding energy can exist in the degenerate semiconductor with the charge-carrier concentration higher than the Mott density[14,15]. Previously, all the optical absorption spectra of $Bi_4O_4SeCl_2$ were measured in the energy range above ~ 500 meV[25,26]. However, excitons have rarely been observed in $Bi_4O_4SeCl_2$ or even its parent compounds—$Bi_2O_2Se$ and $Bi_2O_2Cl_2$ (or the counterpart $Bi_2O_2Br_2$)[25,28,29]. Infrared spectroscopy is an efficient experimental technique for investigating low-energy excitations of materials in the photon energy range at least down to 10 meV and thus sheds light on seeking excitons—one type of elementary excitations (i.e., quasi-particles) in $Bi_4O_4SeCl_2$[30-37].

Here, we report a combined infrared spectroscopy, electrical transport, ab initio calculation, and angle-resolved-photoemission spectroscopy study of a van der Waals degenerate-semiconductor $Bi_4O_4SeCl_2$. A peak-like feature (i.e., $\alpha$ peak) can be observed around ~ 125 meV in its optical conductivity spectra at not only low temperature $T$ = 8 K but also room temperature. Considering that the $\alpha$ peak is irrelevant with a Drude component, interband transitions, the localized states and the polaron absorption, we attribute the $\alpha$ peak to the optical absorption of the room-temperature-stable excitons in $Bi_4O_4SeCl_2$. Besides, if the room-temperature-stable excitons in $Bi_4O_4SeCl_2$ were weakly-bound, the Lyddane-Sachs-Teller relation would be violated in this material. Additionally, the exciton binding energy of ~ 375 meV, which is about an order of magnitude larger than those of conventional semiconductors, and the charge-carrier concentration of ~ $1.25 \times 10^{19}$ cm$^{-3}$, which is higher than the Mott density, further indicate that the excitons in this highly-doped van der Waals degenerate semiconductor should be tightly bound.

## Results and discussion

**Crystal structure characterization.** Before performing the infrared spectroscopy measurements, we grew its single crystals using a modified Bridgman method (see the details about the growth of the single crystals in the Methods Section)[25]. Since X-ray diffraction (XRD) and transmission electron microscopy (TEM) can provide crucial information on crystalline structures, the XRD and TEM measurements were performed to check whether the samples synthesized by us are the $Bi_4O_4SeCl_2$ single crystals. The Rietveld refinement of the powder XRD pattern of our samples in Fig. 1a shows the lattice parameters $a = b = 3.908$ Å and $c = 27.066$ Å, which are consistent with the previously reported lattice parameters of $Bi_4O_4SeCl_2$ (see the schematic drawing of the crystal structure in the inset of Fig. 1a) [25]. In addition, we performed the Rietveld refinement with Se/Cl mixing on the two anion sites and obtained the goodness of fit parameter $R_p \sim 6.52\%$, the weighted profile $R$-factor $R_{wp} \sim 9.13\%$, and the statistical parameter evaluating the overall agreement between the measured and calculated XRD data $\chi^2 = 7.35$ (see Supplementary Fig. S1a). Furthermore, the Rietveld refinement based on the crystal structure of $Bi_4O_4SeCl_2$ in the inset of Fig. 1a yields the parameters $R_p \sim 5.98\%$, $R_{wp} \sim 8.25\%$, and $\chi^2 = 6.02$. Compared with the Rietveld refinement with Se/Cl mixing on the two anion sites, the Rietveld refinement based on the crystal structure in the inset of Fig. 1a has the smaller values of $R_p$, $R_{wp}$ and $\chi^2$, which suggests a smaller difference between the crystal structure in Fig. 1a and the crystal structure corresponding to the measured XRD data. In fact, the Rietveld refinement based on the crystal structure in Fig. 1a shows that the O and Se atoms have the vacancy rates of ~ 0.3, respectively, which suggests that vacancy defects should be un-negligible in our $Bi_4O_4SeCl_2$ single crystals (See Supplementary Fig. S1b). Moreover, as displayed in Fig. 1b, only (00$l$) peaks were present in the diffraction XRD data of our single crystals, which indicates that the cleaved single-crystal surface is perpendicular to the crystallographic $c$-axis (i.e., the cleaved surface is the $ab$-plane). The high-angle annular dark-field image (HADDF) in Fig. 1c, which was obtained by the TEM measurements of our single crystals, shows the atomic distributions of Bi (or Se) along the [001] zone axis in real space (see the brightest star-like pattern in the top-left inset of Fig. 1c). Thus, the XRD and TEM measurements confirmed that the grown samples here are the $Bi_4O_4SeCl_2$ single crystals.

**Infrared spectroscopy and charge transport measurements.** Using infrared spectroscopy, we measured the optical reflectance spectra $R(\omega)$ of the $Bi_4O_4SeCl_2$ single crystals at different temperatures with the electrical field of the incident light applied parallel to the $ab$-plane over a broad energy range (see the details about the reflectance measurements in the Methods Section). Fig. 2a presents the $R(\omega)$ of the $Bi_4O_4SeCl_2$ single crystals in the energy range up to 600 meV at two representative temperatures $T = 8$ K and 300 K (see the $R(\omega)$ up to 3.4 eV in the inset of Fig. 2a and the $R(\omega)$ of the $Bi_4O_4SeCl_2$ single crystal from a different growth batch at two typical temperatures $T = 8$ K and 300 K in Supplementary Fig. S2a). To further identify the optical signatures of low-energy excitations, we obtained the real part (i.e., $\sigma_1(\omega)$) of the *ab*-plane optical conductivity of $Bi_4O_4SeCl_2$ via the Kramers-Kronig transformation of the measured $R(\omega)$. The $\sigma_1(\omega)$ of the $Bi_4O_4SeCl_2$ single crystals at $T = 8$ K and 300 K were plotted in Fig. 2b. A Drude component—upturn-like feature arising from the optical response of the free carriers can be observed at energies lower than 75 meV in the $\sigma_1(\omega)$, which is consistent with the metallic behavior of the temperature-dependent resistivity (i.e., decreasing with decreasing temperature) in Fig. 2c (see the details about the electrical transport measurement in the Methods Section) and is also present in the $\sigma_1(\omega)$ of the $Bi_4O_4SeCl_2$ single crystal from the different growth batch (see Supplementary Fig. S2b). However, as a physical quantity proportional to the charge-carrier concentration, the spectral weight ($\sim 1.9 \times 10^5\ \Omega^{-1}\ cm^{-2}$) of the Drude component of the $Bi_4O_4SeCl_2$ single crystals is about three orders of magnitude lower than those of good metals (see the Drude-Lorentz fit to the $\sigma_1(\omega)$ of the $Bi_4O_4SeCl_2$ single crystal from a different growth batch at $T = 8$ K in Supplementary Fig. S2c and the obtained Drude weight of $\sim 1.5 \times 10^5\ \Omega^{-1}\ cm^{-2}$), such as gold with the Drude weight of $\sim 1.2 \times 10^8\ \Omega^{-1}\ cm^{-2}$ [38], which suggests that the charge-carrier concentration in the $Bi_4O_4SeCl_2$ single crystals is about three orders of magnitude smaller than those of good metals, e.g., the bulk gold has the carrier concentration of $\sim 5.91 \times 10^{22}\ cm^{-3}$ [38]. The Hall resistivity $\rho_{xy}$ of the $Bi_4O_4SeCl_2$ single crystals in Fig. 2d scales linearly with the magnetic field applied perpendicular to the crystalline *ab*-plane (see Supplementary Fig. S3 and the details about electrical transport measurements in the Methods Section). According to the inverse linear relationship between the Hall conductivity $\rho_{xy}$ and the charge-carrier density $n$, we obtained the charge-carrier density at each temperature $T$. The charge-carrier densities were plotted as a function of temperature in the inset of Fig. 2d. In Fig. 2d, the charge-carrier density shows a quite weak temperature dependence and has the value of $\sim (1.25 \pm 0.05) \times 10^{19}\ cm^{-3}$, which is indeed three orders of magnitude lower than that of gold. This carrier concentration in $Bi_4O_4SeCl_2$ implies a significantly reduced Coulomb screening and thus provides a chance to explore quasi-particles including excitons in this van der Waals material.

**Peak-like feature around ~ 125 meV.** It is worth noticing that a peak-like feature $\alpha$ is present around ~ 125 meV in the $\sigma_1(\omega)$ of the $Bi_4O_4SeCl_2$ single crystals, as indicated by a red arrow in Fig. 2b. This peak-like feature around ~ 125 meV can be also observed in the $\sigma_1(\omega)$ of the $Bi_4O_4SeCl_2$ single crystal from the different growth batch (see Supplementary Fig. S2b). As illustrated in the schematic drawing of the $\sigma_1(\omega)$ of a material (see Fig. 3a), the origin of this $\alpha$ peak remains elusive because (i) considering that this $\alpha$ peak is close to the low-energy Drude component, the $\alpha$ peak may correspond to another Drude component[39]; (ii) since the $\alpha$ peak is mainly located at higher energies than the low-energy Drude component, the $\alpha$ peak is likely to arise from interband transitions[25-27]; (iii) the $\alpha$ peak may be associated with the localized states coming from impuries and defects (see the schematic drawing of the localized states in the inset of Fig. 3a) [40]; (iv) given that the $\alpha$ peak is located above the phonon at ~ 37 meV, the $\alpha$ peak is likely to be contributed by the polaron absorption[41-43]; (v) since the $\alpha$ peak is located below the bandgap of ~ 500 meV extracted from by the linear extrapolation of the $\sigma_1(\omega)$ (see the black dashed lines in the inset of Fig. 2b and the similar bandgap of the $Bi_4O_4SeCl_2$ single crystal from the different growth batch in Supplementary Fig. S2b) [44], and excitons can be present in degenerate semiconductors[14-16], the $\alpha$ peak may originate from the exciton absorption in this degenerate semiconductor. Therein, as shown by the gray curve in Fig. 3b, the fit to the low-energy part of the $\sigma_1(\omega, T = 8\text{ K})$ at energies below 300 meV using the two Drude components of a standard Drude-Lorentz model cannot reproduce the $\alpha$ peak (see the details about the Drude-Lorentz model in the Methods Section and the fitting parameters in Table 1), which indicates that the $\alpha$ peak should not arise from the optical response of the free carriers. In contrast, Fig. 3c displays that the low-energy part of the $\sigma_1(\omega, T = 8\text{ K})$ at $\omega \leq 300$ meV can be well fit by the Drude component and the Lorentzian peak of the Drude-Lorentz model (see the fitting parameters in Table 2 and the fitting curve at higher energies in Supplementary Fig. S4), i.e., the $\alpha$ peak around ~ 125 meV in the $\sigma_1(\omega, T = 8\text{ K})$ can be reproduced using a Lorentzian peak. Besides, the $\alpha$ peak in the $\sigma_1(\omega, T = 300\text{ K})$ can also be well fit by a Lorentzian peak (see Fig. 3d, the fitting parameters in Table 3 and the fitting curve at higher energies in Supplementary Fig. S5). To further investigate the origin of this $\alpha$ peak around ~ 125 meV in the $\sigma_1(\omega)$ the $Bi_4O_4SeCl_2$ single crystals, it is essential to obtain the electronic band structure and lattice vibration modes of $Bi_4O_4SeCl_2$ because the possible mechanisms for the presence of the $\alpha$ peak are related to interband transitions, localized states, polarons or excitons.

Figures 4a and 4b show the theoretical and experimental electronic band structures of $Bi_4O_4SeCl_2$, which were gotten using first-principle calculations and angle-resolved photoemission spectroscopy (APRES), respectively[45-47]. Comparing the Fermi levels of the electronic bands around the Γ point obtained by first-principle calculations and ARPES measurements, the Fermi level of the calculated electronic bands should be moved up by ~ 410 meV (see the experimental Fermi level represented by the red dashed horizontal line in Fig. 4a). According to the Pauli exclusion principle and the Fermi level obtained by the ARPES measurements, the original *indirect* optical transitions from the valence-band maxima close to the X and R points to the conduction-band minimum at the Γ point are forbidden (see the black dashed arrows in Fig. 4a), while the *direct* interband transitions with the energy scale of ~ 500 meV at the Γ point are allowed (see the red arrow in Fig. 4a). Since the allowed *direct* interband transitions at the Γ point have the energy scale (~ 500 meV) much larger than that of the $\alpha$ peak, the $\alpha$ peak should be irrelevant with the interband transitions. Furthermore, as shown in the inset of Fig. 3a, the localized states coming from impurities and defects may be present within the bandgap. Given the Fermi energy (~ 0.17 eV) obtained by the ARPES measurements and the Pauli exclusion principle, the minimal energy for the optical transition from the occupied localized states to the empty conduction band above the Fermi level is ~ 0.17 eV (see the arrows denoting the allowed optical transition related to the localized states in the inset of Fig. 3a), which is also higher than the $\alpha$ peak energy. Thus, the $\alpha$ peak is very unlikely to arise from the optical transitions from the localized states to the conduction bands.

Conventional polarons can be formed due to the interactions between electrons (or holes) with longitudinal-optical (LO) phonons[48]. To check whether the presence of the $\alpha$ peak is significantly contributed by the polaron absorption, we performed first-principle calculations of the phonon modes in $Bi_4O_4SeCl_2$ (see the details about the first-principle calculations in the Methods Section). Table 4 shows the energies and species of the calculated phonon modes. Comparing the energies of the calculated phonon modes and the peak-like features in the $\sigma_1(\omega)$ of $Bi_4O_4SeCl_2$, the sharp peak-like feature present around ~ 34.6 meV in the $\sigma_1(\omega)$ can be assigned as the longitudinal infrared-active phonon mode with the energy of ~ 38.8 meV. Moreover, the optical absorption feature of the large polaron (i.e., Fröhlich polaron) is usually present at higher energy than that of the LO phonon, which is consistent with the case that the $\alpha$ peak around 125 meV here is located at higher energy than the LO phonon at ~ 34.6 meV and thus indicates that it is necessary to study whether the $\alpha$ peak results from the large polaron relevant with the observed LO infrared-active phonon. It was reported that when the plasma energy $\omega_p$ is lower than (or equal to) the LO-phonon energy, the LO phonon would be insufficiently screened by charge carriers, which leads to the observation of the polaron feature[42,43]. However, as listed in Table 2, the plasma energies of $Bi_4O_4SeCl_2$ (i.e., $\omega_p(T = 8\text{ K}) \approx 334$ meV and $\omega_p(T = 300\text{ K}) \approx 314$ meV) are much higher than the LO-phonon energy $\omega_{LO} \approx$ 34.6 meV, which indicates that the LO phonon with the energy of ~ 34.6 meV is significantly screened by the charge carriers in $Bi_4O_4SeCl_2$ and therefore suggests that the $\alpha$ peak should not originate from the optical absorption of the large polaron.

Accompanied with the occurrence of optical interband transitions, excitonic states can form and exist within the bandgap[1-4]. A natural question to ask is which interband transition the excitons in $Bi_4O_4SeCl_2$ correspond to. According to the Pauli exclusion principle and the measured Fermi level, the direct interband transitions with the energy scale of ~ 500 meV at the Γ point (see the red arrow in Fig. 4a) take place. Moreover, the energy scale (~ 500 meV) of the direct interband transitions indicated by the red arrow in Fig. 4a is consistent with the bandgap (~ 500 meV) obtained by the linear extrapolation of the $\sigma_1(\omega)$ (see the black dashed line in the inset of Fig. 2b). Thus, the $\alpha$ peak below the bandgap can be assigned as the exciton absorption accompanied with the direct interband transitions at the Γ point. Moreover, we notice that the edge-like absorption feature arising from the indirect interband transition was previously reported to be at the energy near 1 eV[25,26], which seems to be absent in the inset of Fig. 2b. In order to check the existence of the indirect-interband-transition-induced absorption feature at the energy near 1 eV, we fit the $\sigma_1(\omega)$ of our $Bi_4O_4SeCl_2$ single crystals using a standard Drude-Lorentz model (see the Drude and Lorentzian components in Fig. S4 of Supplemental Information and the fitting parameters in Table 2). Then, we removed the direct-interband-transition-induced Lorentzian component present in the energy range from ~ 0.5 to ~ 1.2 eV (see the residual $\sigma_1(\omega)$ and the fitting parameters in Supplementary Table 1S). Finally, an edge-like feature with the energy intercept of ~ 1 eV under the linear extrapolation can be observed in the absorption coefficient $\alpha(\omega)$ (see Supplementary Fig. S6b). Therefore, the *direct*-interband-transition-induced edge-like absorption feature makes the edge-like absorption feature caused by the *indirect* interband transition with energy scale of ~ 1 eV seem absent.

**Tightly-bound excitons.** To check whether the excitons in $Bi_4O_4SeCl_2$ can be regarded as tightly-bound excitons or not, we first studied the static dielectric constant of the assumed weakly-bound excitons—Wannier-type excitons in this system. According to the effective Rydberg energy of Wannier-type excitons $R^* = \mu R_H/(m_0\varepsilon_r^2)$, where $\mu$ is the reduced mass of the excitons, the Rydberg energy $R_H$ = 13.6 eV, and $m_0$ is the bare electron mass, we can obtain the static dielectric constant $\varepsilon_r$. Here, (i) the Rydberg energy is equal to the difference between the bandgap and the 1st-order-exciton absorption energy, i.e., $R^*$ = 375 meV, (ii) the reduced mass $\mu$ of the excitons has the relationship with the effective electron mass $m_{el}$ and the effective hole mass $m_h$, $1/\mu = 1/m_{el} + 1/m_h$, so $\mu = 0.139\ m_0$ when both the effective electron mass $m_{el}$ and the effective hole mass $m_h$ are assumed to be equal to the conduction band electrons $m^*$ measured by ARPES, i.e., $m_{el} = m_h = m^* = 0.278\ m_0$ (here, the energy dispersion of the measured conduction band can be obtained by fitting the trace using a parabola, i.e., $E(k) = -\ 0.17 + 13.7\ k^2$, where the energy unit is eV, and the momentum unit is $Å^{-1}$. According to the relationship that the curvature of a band is inversely proportional to the effective mass, the conduction band effective mass $m^*$ shown in our manuscript was estimated to be 0.278 $m_0$. To obtain clearer band dispersions, we further performed surface potassium-dosing to raise the carrier density of the $Bi_4O_4SeCl_2$ single crystals and measured the band structure of the $Bi_4O_4SeCl_2$ single crystal with a higher carrier density using ARPES. As displayed in Supplementary Fig. S7, after surface potassium-dosing, the ARPES-derived conduction band of the $Bi_4O_4SeCl_2$ single crystal with a clearer energy dispersion exhibits a rigid band shift and has a similar effective mass $m^* = 0.278\ m_0$). Therefore, we can obtain the static dielectric constant of the assumed Wannier-type excitons $\varepsilon_r \approx 2.25$. Here, the estimated $\varepsilon_r$ is distinctly smaller than the static dielectric constant (the order of ~ 10) of Wannier-type excitons. According to the Lyddane-Sachs-Teller relation with the static dielectric constant larger than the infinite dielectric constant[49], infinite dielectric constant gives a lower limit of the value of static dielectric constant quantifying the dielectric screening of the Coulomb interactions between the electron and hole of an exciton and therefore can provide important information on the nature of an exciton. Here, the estimated $\varepsilon_r$ is small than the infinite dielectric constant $\varepsilon_\infty \approx 5.94$ (see the details about the estimation of the infinite dielectric constant $\varepsilon_\infty$ in the Methods Section), which violates the Lyddane-Sachs-Teller relation. Thus, considering the relatively small value of the static dielectric constant of the assumed Wannier-type excitons, the excitons in $Bi_4O_4SeCl_2$ should not be weakly bound. Moreover, as a physical quantity reflecting the strength of the electron-hole interactions in excitons, the exciton binding energy $E_b$ of $Bi_4O_4SeCl_2$ can be obtained by subtracting the $\alpha$ peak energy from the bandgap, i.e., $E_b \approx 375$ meV. Here, the exciton binding energy $E_b$ of $Bi_4O_4SeCl_2$ is about an order of magnitude larger than those of most conventional semiconductors hosting weakly-bound excitons, which suggests the tightly-bound excitons in $Bi_4O_4SeCl_2$. It is worth noticing that the exciton binding energy is over one order of magnitude higher than the thermal energy of room temperature, which is consistent with the observation of the exciton absorption feature—$\alpha$ peak at $T = 300$ K.

To examine the existence of tightly-bound excitons in $Bi_4O_4SeCl_2$, we compared the structural characteristic of this van der Waals material with those of several typical transition metal dichalcogenides hosting tightly-bound excitons, which include the $WSe_2$ monolayers, the $WS_2$ monolayers, and the $WS_2$ bulk crystals. The two adjacent quasi-2D $WSe_2$ layers and the two adjacent quasi-2D $WS_2$ layers have the distances of ~ 6.49 Å and ~ 6.18 Å, respectively (see Supplementary Fig. S8)[50,51]. The relatively large distances between the quasi-2D adjacent layers in $WSe_2$ and $WS_2$ were expected to result in the less overlap of the electron clouds along the interlayer directions, which leads to the significant reduction of the Coulomb screening and the formation of tightly bound excitons in $WSe_2$ and $WS_2$[17,52]. It is worth noticing that the distances between the two adjacent Bi-O-Bi layers in $Bi_4O_4SeCl_2$ are ~ 6.23 Å and ~ 7.36 Å, respectively (see Supplementary Fig. S8) [25]. Thus, from the viewpoint of the structural characteristic, the distances between the adjacent Bi-O-Bi layers in $Bi_4O_4SeCl_2$ are comparable to those between the quasi-2D adjacent layers in $WSe_2$ and $WS_2$, which suggests that the reduction of the Coulomb screening in $Bi_4O_4SeCl_2$ may be comparable to those in $WSe_2$ and $WS_2$. In $Bi_4O_4SeCl_2$, although the relatively high charge-carrier density (~ $1.25 \times 10^{19}$ cm$^{-3}$), which is likely to be caused by the presence of impurities and defects probably due to the defect-induced dangling bonds and the difference between the valence electrons of the atoms of $Bi_4O_4SeCl_2$ and the impurity atoms, would result in the enhancement of the static dielectric constant $\varepsilon_r$ and weaken the intensity of the optical absorption feature of excitons, the relatively large distances between its two adjacent Bi-O-Bi layers can lead to the significant reduction of the Coulomb screening, which supports the existence of tightly bound excitons in this van der Waals material. Compared with the tightly bound excitons in the $WSe_2$ monolayers, the $WS_2$ monolayers and the $WS_2$ bulk crystals, the tightly bound excitons in $Bi_4O_4SeCl_2$ have a similar exciton binding energy and indeed exhibit a *weaker* intensity of the optical absorption feature.

To further confirm the tightly-bound excitons in $Bi_4O_4SeCl_2$, we compared the estimated Mott density with the charge-carrier concentration. Before estimating the Mott density, we need to obtain the static dielectric constant $\varepsilon_r = (\mu \cdot 13.6\ \mathrm{eV}/E_\mathrm{b} m_\mathrm{e})^{1/2} \approx 2.25$ and the effective Bohr radius $a_\mathrm{B} = \varepsilon_r a_\mathrm{H} m_e/\mu \approx 8.6$ Å (where $a_\mathrm{H}$ is the Bohr radius). Therefore, the Mott density of $Bi_4O_4SeCl_2$ can be calculated according to the two following formulas[16]:

$$n_M = 1.19^2 \frac{k_\mathrm{B}T}{2a_\mathrm{B}^3 E_\mathrm{b}} = 2.1 \times 10^{18}\ \mathrm{cm}^{-3}\ (T = 8\ \mathrm{K});$$

$$n_M = a_\mathrm{B}^{-3}\ \frac{\pi \cdot 1.19^6}{4^3 \cdot 3} \left(\frac{m_e m_h}{(m_e+m_h)^2}\right)^3 = 1.1 \times 10^{18}\ \mathrm{cm}^{-3}.$$

Here, the charge-carrier concentration $n \sim 1.25 \times 10^{19}$ cm$^{-3}$ is about one order of magnitude higher than the estimated Mott densities, which means that only tightly-bound excitons can survive in $Bi_4O_4SeCl_2$ with the carrier concentration higher than the Mott density.

Given that (i) the larger charge carrier densities of the graphene/ZnO/PbS quantum-dot devices and the $MoS_2$/$MoSe_2$ heterostructure were reported to result in the enhancement of photoresponsivity, due to the increased screening of the interfacial Coulomb potential and the faster transport of charge carriers[24], (ii) the room-temperature stability of excitons is crucial for their wide and realistic applications in optoelectronic devices, such as excitonic light-emitting diodes[1,8,53-55], photodetectors[56], exciton optoelectronic transistors[57], and exciton-based optoelectronic switches[58], (iii) van der Waals materials provide a rich avenue for achieving exceptional phenomena using various methods, including applying gate voltage and reducing the material thickness[59], and (iv) tightly-bound excitons were revealed to exist at room temperature in van der Waals degenerate-semiconductor $Bi_4O_4SeCl_2$ with high charge-carrier density, our study provides a new material candidate for the development of the optoelectronic devices based on the tightly-bound and room-temperature-stable excitons in highly-doped van der Waals degenerate semiconductors.

## Conclusion

In summary, using infrared spectroscopy, electrical transport, first-principle calculations, and angle-resolved-photoemission spectroscopy, we have investigated the nature of the peak-like feature (i.e., $\alpha$ peak) present around ~ 125 meV in the $\sigma_1(\omega)$ of a van der Waals degenerate semiconductor $Bi_4O_4SeCl_2$ with a high carrier concentration at temperatures $T$ = 8 K and 300 K. Because the $\alpha$ peak cannot be reproduced using a Drude component, the $\alpha$ peak is unlikely to arise from the optical response of free charge carriers. Moreover, the allowed interband transition energy (> 500 meV) is much higher than the $\alpha$ peak energy, so the $\alpha$ peak should be irrelevant with to the interband transitions. Besides, the minimal energy (~ 170 meV) for the optical transition related to the localized states is distinctly larger than the $\alpha$ peak energy, so the $\alpha$ peak should not be associated with the localized states. In addition, since the plasma energies (~ 334 meV at 8 K and ~ 314 meV at 300 K) are much higher than the calculated LO-phonon energy of ~ 34.6 meV, the $\alpha$ peak should not originate from the optical absorption of the polaron. Considering that the $\alpha$ peak should be unrelated to a Drude component, interband transitions, the localized states and the polaron absorption, we attributed the $\alpha$ peak to the optical absorption of the excitons which are stable at room temperature in $Bi_4O_4SeCl_2$. Furthermore, if the excitons in this material were assumed to be weakly-bound excitons—Wannier-type excitons, the Lyddane-Sachs-Teller relation would be violated. Besides, the high exciton binding energy of ~ 375 meV which is about an order of magnitude larger than those of conventional semiconductors, combined with the large charge-carrier concentration of ~ $1.25 \times 10^{19}$ cm$^{-3}$ which is higher than the Mott density, further indicates that the excitons in this highly-doped system should be tightly bound. Our work provides a new option for the development of the optoelectronic devices based on the tightly-bound and room-temperature-stable excitons in highly-doped van der Waals degenerate semiconductors.

## Methods

**Sample preparation.** The $Bi_4O_4SeCl_2$ single crystals were synthesized from a stoichiometric mixture of BiOCl, Bi, $Bi_2O_3$, and Se powder sealed in an evacuated quartz tube. The quartz tube was gradually heated up to 800°C and maintained for 24 h. Afterward, the quartz tube slowly heated up to 1180°C and kept for 24 h. Then, the quartz tube slowly cooled to 900°C over 120 h and finally cooled to room temperature. The plate-like single crystals of $Bi_4O_4SeCl_2$ were obtained.

**Optical reflectance measurements.** The reflectance spectra $R(\omega)$ was measured over a broad energy range using a Bruker Vertex 80v Fourier transform spectrometer. An optical cone was used to mount the single-crystal sample of $Bi_4O_4SeCl_2$ at the cold finger of a heliun flow cryostat. An in situ gold and aluminum overcoating technique was used to get the reflectance spectra $R(\omega)$. All optical reflectance measurements were performed on the freshly cleaved (001) surfaces.

**Electrical transport measurements.** Both longitudinal and Hall electrical resistivity measurements were performed in four-probe configuration with the electric currents applied in (001) surfaces of rectangular shape $Bi_4O_4SeCl_2$ single crystals. Moreover, the magnetic field for the measurement of the Hall electrical resistivity was applied perpendicular to the (001) surfaces of $Bi_4O_4SeCl_2$ single crystals.

**Kramers-Kronig transformation** The real parts of optical conductivity $\sigma_1(\omega)$ were obtained by the Kramers-Kronig transformation of the reflectance spectra $R(\omega)$. The phase shift (i.e., $\theta(\omega)$) of the reflected light relative to the incident light was determined using the Kramers-Kronig transformation of the reflectance spectra $R(\omega)$ of $Bi_4O_4SeCl_2$. The $R(\omega)$ were extrapolated by the Hagen-Rubens relation in the low-energy side. The relationship between optical constants can be used to determine the real part $\sigma_1(\omega)$ and the imaginary part $\sigma_2(\omega)$ of the optical conductivity when the $R(\omega)$ and $\theta(\omega)$ are known.

**Drude-Lorentz model.** The standard Drude-Lorentz Model has the following form:

$$\sigma_1(\omega)=\sum_{j=1}^{N}\frac{2\pi}{Z_0}\frac{\omega_{\mathrm{p_j}}^2\Gamma_{\mathrm{D_j}}}{\omega^2+\Gamma_{\mathrm{D_j}}^2}+\sum_{j=1}^{N}\frac{2\pi}{Z_0}\frac{S_j^2\omega^2\Gamma_j}{\left(\omega_j^2-\omega^2\right)^2+\omega^2\Gamma_j^2}, \quad (1)$$

where $Z_0 \approx 377\ \Omega$ is the impedance of free space, $\omega_{\mathrm{p}j}$ is the plasma energy, and $\Gamma_{\mathrm{D}j}$ is the relaxation rate of free charge carriers, while $\omega_j$, $\Gamma_j$, and $S_j$ are the resonance energy, the damping, and the mode strength of each Lorentz term, respectively. The first term in Equation (1) denotes the optical response of free carriers, i.e., Drude component. The Lorentzian peak can be described by the second term in Equation (1).

**First-principle calculations.** The first-principle calculations on $Bi_4O_4SeCl_2$ were carried out using the Vienna ab initio simulation package with the generalized gradient approximation (GGA) of the Perdew-Burke-Ernzerhof (PBE) exchange correlation potential[60,61]. The plane-wave cutoff energy 500 eV and the 13 × 13 × 2 *k*-point mesh were utilized for the calculations of band structure. The spin-orbit coupling (SOC) effects were taken into account in our calculations. To evaluate the phonon frequencies and the corresponding phonon vibrational modes of $Bi_4O_4SeCl_2$, the density-functional-perturbation theory method implemented in phonopy package was used[62,63]. The 3 × 3 × 1 supercell with 198 atoms is used in the phonon calculation. Before the phonon calculation, the crystal structure with the initial lattice parameters of $a = b = 3.908$ Å and $c = 27.066$ Å are fully relaxed with the forces and energy convergence of 0.01 eV $Å^{-1}$ and $1 \times 10^{-5}$ eV, respectively. In order to accurately describe the interlayer van der Waals interactions, the vdW density functional method of the SCAN+rVV10 together with the spin-orbit coupling effect are adopted as described in the previous paper[25,64]. A plane-wave kinetic energy cutoff of 550 eV is set in phonon calculations. The Monkhorst Pack *k*-point mesh of 7 × 7 × 1 and 5 × 5 × 3 are used for the structural optimization and phonon properties calculations, respectively.

**ARPES experiments.** The ARPES experiments were carried out on the lab-based ARPES using photon energy of 21.2 eV. By cleaving single crystal samples in-situ in a vacuum greater than $5\times10^{-11}$ Torr, clean (001) surfaces of $Bi_4O_4SeCl_2$ were obtained. Data were recorded by a Scienta Omicron DA30L analyser at the sample temperature below 10 K with an overall energy and angle resolutions of 20 meV and 0.2°, respectively.

**Estimating the infinite dielectric constant.** The infinite dielectric constant $\varepsilon_\infty$ can be determined by $\omega_p^* = \omega_p/\sqrt{\varepsilon_\infty}$. We obtained the unscreened plasma frequency $\omega_p \approx 334$ meV from the Drude spectral weight (see Fig. 3c and Table 2). In addition, the screened plasma frequency $\omega_p^*$ is equal to the energy at which the real part of the dielectric function $\varepsilon_1(\omega) = 0$, i.e., $\omega_p^* \approx 137$ meV (see Supplementary Fig. S9). Thus, the infinite dielectric constant $\varepsilon_\infty \approx 5.94$.

## Data availability

The data that support the findings of this study are available from the corresponding author upon reasonable request.

## Additional information

**Supporting information**. The online version contains supplementary material available.

## Author contributions

Z.-G.C., J.-G.G. and N.X. conceived and supervised this project. Y.X. carried out the optical reflectance experiments. J.W. grew the single crystals, performed the electrical transport and XRD measurements. B.S. did the calculations of the phonon modes. J.D. performed the calculations of the electronic band structure. C.P. and C.W. did the ARPES measurements. Q.Z. and L.G. performed the TEM measurements. Z.-G.C., J.-G.G., N.X., J. L. and Y.X. analyzed the data. Z.-G.C. wrote the paper with inputs from all authors.

## Competing Interests

The authors declare no conflict of interest.

## References

1. Mak, K. F. & Shan, J. Photonics and optoelectronics of 2D semiconductor transition metal dichalcogenides. *Nat. Photonics* **10**, 216-226 (2016).
2. Wang, G. et al. Colloquium: Excitons in atomically thin transition metal dichalcogenides. *Rev. Mod. Phys.* **90**, 021001 (2018).
3. Jiang, Y., Wang, X. & Pan, A. Properties of Excitons and Photogenerated Charge Carriers in Metal Halide Perovskites. *Adv. Mater.* **31**, 1806671 (2019).
4. Regan, E. C. et al. Emerging exciton physics in transition metal dichalcogenide heterobilayers. *Nat. Rev. Mater.* **7**, 778-795 (2022).
5. Kogar, A. et al. Signatures of exciton condensation in a transition metal dichalcogenide. *Science* **358**, 1314-1317 (2017).
6. Wang, Z. et al. Evidence of high-temperature exciton condensation in two-dimensional atomic double layers. *Nature* **574**, 76-80 (2019).
7. Su, J.-J. & MacDonald, A. H. How to make a bilayer exciton condensate flow. *Nat. Phys.* **4**, 799-802 (2008).
8. Ross, J. S. et al. Electrically tunable excitonic light-emitting diodes based on monolayer $WSe_2$ p–n junctions. *Nat. Nanotechnol.* **9**, 268-272 (2014).
9. Ma, Z. et al. High Color-Rendering Index and Stable White Light-Emitting Diodes by Assembling Two Broadband Emissive Self-Trapped Excitons. *Adv. Mater.* **33**, 2001367 (2021).
10. Wang, J. et al. Polarized Light-Emitting Diodes Based on Anisotropic Excitons in Few-Layer $ReS_2$. *Adv. Mater.* **32**, 2001890 (2020).
11. Moskalenko, S. A. e., Moskalenko, S. & Snoke, D. *Bose-Einstein condensation of excitons and biexcitons: and coherent nonlinear optics with excitons*. Cambridge University Press (2000).
12. Mott, N. F. Metal-Insulator Transition. *Rev. Mod. Phys.* **40**, 677-683 (1968).

13. Klingshirn, C. F. *Semiconductor optics*. Springer Science & Business Media (2012).
14. Mahan, G. D. Excitons in degenerate semiconductors. *Phys. Rev.* **153**, 882-889 (1967).
15. Mahan, G. D. Excitons in Metals. *Phys. Rev. Lett.* **18**, 448-450 (1967).
16. Palmieri, T. et al. Mahan excitons in room-temperature methylammonium lead bromide perovskites. *Nat. Commun.* **11**, 850 (2020).
17. He, K. et al. Tightly bound excitons in monolayer $WSe_2$. *Phys. Rev. Lett.* **113**, 026803 (2014).
18. Calman, E. V. et al. Indirect excitons in van der Waals heterostructures at room temperature. *Nat. Commun.* **9**, 1895 (2018).
19. Zhang, K. et al. Room-Temperature Magnetic Field Effect on Excitonic Photoluminescence in Perovskite Nanocrystals. *Adv. Mater.* **33**, 2008225 (2021).
20. Liu, Y. et al. Interlayer Excitons in Transition Metal Dichalcogenide Semiconductors for 2D Optoelectronics. *Adv. Mater.* **34**, 2107138 (2022).
21. Geim, A. K. & Grigorieva, I. V. Van der Waals heterostructures. *Nature* **499**, 419-425 (2013).
22. Ajayan, P., Kim, P. & Banerjee, K. Two-dimensional van der Waals materials. *Phys. Today* **69**, 38-44 (2016).
23. Basov, D. N., Fogler, M. M. & García de Abajo, F. J. Polaritons in van der Waals materials. *Science* **354**, aag1992 (2016).
24. Wang, H. et al. Van der Waals Integration Based on Two-Dimensional Materials for High-Performance Infrared Photodetectors. *Adv. Funct. Mater.* **31**, 2103106 (2021).
25. Gibson, Q. D. et al. Modular design via multiple anion chemistry of the high mobility van der Waals semiconductor $Bi_4O_4SeCl_2$. *J. Am. Chem. Soc.* **142**, 847-856 (2019).
26. Ji, R. et al. Multiple anion chemistry for ionic layer thickness tailoring in $Bi_{2+2n}O_{2+2n}Se_nX_2$ (X = Cl, Br) van der Waals semiconductors with low thermal conductivities. *Chem. Mater.* **34**, 4751-4764 (2022).
27. Newnham, J. A. et al. Band Structure Engineering of $Bi_4O_4SeCl_2$ for Thermoelectric Applications. *ACS Org. Inorg. Au* **2**, 405-414 (2022).
28. Wu, J. et al. High electron mobility and quantum oscillations in non-encapsulated ultrathin semiconducting $Bi_2O_2Se$. *Nat. Nanotechnol.* **12**, 530-534 (2017).

29. Li, H. et al. Oxygen Vacancy-Mediated Photocatalysis of BiOCl: Reactivity, Selectivity, and Perspectives. *Angew. Chem. Int. Ed.* **57**, 122-138 (2018).
30. Basov, D. N. & Timusk, T. Electrodynamics of high-$T_c$ superconductors. *Rev. Mod. Phys.* **77**, 721-779 (2005).
31. Basov, D. N. et al. Electrodynamics of correlated electron materials. *Rev. Mod. Phys.* **83**, 471-541 (2011).
32. Dressel, M. & Grüner, G. *Electrodynamics of solids: Optical properties of electrons in matter*. Cambridge University Press (2002).
33. Kim, S. Y. et al. Spectroscopic studies on the metal-Insulator transition mechanism in correlated materials. *Adv. Mater.* **30**, 1704777 (2018).
34. Chen, Z. G. et al. Measurement of the *c*-axis optical reflectance of $AFe_2As_2$ (A=Ba, Sr) single crystals: evidence of different mechanisms for the formation of two energy gaps. *Phys. Rev. Lett.* **105**, 097003 (2010).
35. Dai, Y. M. et al. Hidden T-linear scattering rate in $Ba_{0.6}K_{0.4}Fe_2As_2$ revealed by optical spectroscopy. *Phys. Rev. Lett.* **111**, 117001 (2013).
36. Xu, B. et al. Temperature-driven topological phase transition and intermediate Dirac semimetal phase in $ZrTe_5$. *Phys. Rev. Lett.* **121**, 187401 (2018).
37. Su, B. et al. Strong and tunable electrical anisotropy in type-II Weyl semimetal candidate $WP_2$ with broken inversion symmetry. *Adv. Mater.* **31**, 1903498 (2019).
38. Olmon, R. L. et al. Optical dielectric function of gold. *Phys. Rev. B* **86**, 235147 (2012).
39. Tu, J. J. et al. Optical properties of the iron arsenic superconductor $BaFe_{1.85}Co_{0.15}As_2$. *Phys. Rev. B* **82**, 174509 (2010).
40. Dumke, W. P. Optical Transitions Involving Impurities in Semiconductors. *Phys. Rev.* **132**, 1998-2002 (1963).
41. Alexandrov, A. S. & Devreese, J. T. *Advances in polaron physics*. Springer (2010).
42. Devreese, J. T. & Alexandrov, A. S. Fröhlich polaron and bipolaron: recent developments. *Rep. Prog. Phys.* **72**, 066501 (2009).
43. Fujioka, J., Yamada, R., Okawa, T. & Tokura, Y. Dirac polaron dynamics in the correlated semimetal of perovskite $CaIrO_3$. *Phys. Rev. B* **103**, L041109 (2021).

44. Chen, Z. G. et al. Infrared spectrum and its implications for the electronic structure of the semiconducting iron selenide $K_{0.83}Fe_{1.53}Se_2$. *Phys. Rev. B* **83**, 220507(R) (2011).
45. Kotliar, G. et al. Electronic structure calculations with dynamical mean-field theory. *Rev. Mod. Phys.* **78**, 865-951 (2006).
46. Sato, K. et al. First-principles theory of dilute magnetic semiconductors. *Rev. Mod. Phys.* **82**, 1633-1690 (2010).
47. Sobota, J. A., He, Y. & Shen, Z.-X. Angle-resolved photoemission studies of quantum materials. *Rev. Mod. Phys.* **93**, 025006 (2021).
48. Landau, L. D. Electron motion in crystal lattices. *Phys. Z. Sowjet.* **3**, 664 (1933).
49. Lyddane, R. H., Sachs, R. G. & Teller, E. On the Polar Vibrations of Alkali Halides. *Phys. Rev.* **59**, 673-676 (1941).
50. Kulikov, L. et al. Intercalation of niobium, molybdenum, and tungsten diselenides by copper, zinc, and gallium. *Inorg. Mater.* **28**, 397-401 (1992).
51. Gardinier, C. F. & Chang, L. L. Y. Phase relationships in the systems Mo-Sn-S, W-Sn-S and Mo-W-S. *J. Less-Common Met.* **61**, 221-229 (1978).
52. Chernikov, A. et al. Exciton binding energy and nonhydrogenic Rydberg series in monolayer $WS_2$. *Phys. Rev. Lett.* **113**, 076802 (2014).
53. Cheng, R. et al. Electroluminescence and Photocurrent Generation from Atomically Sharp $WSe_2$/$MoS_2$ Heterojunction p–n Diodes. *Nano Lett.* **14**, 5590-5597 (2014).
54. Pospischil, A., Furchi, M. M. & Mueller, T. Solar-energy conversion and light emission in an atomic monolayer p–n diode. *Nat. Nanotechnol.* **9**, 257-261 (2014).
55. Sundaram, R. S. et al. Electroluminescence in Single Layer $MoS_2$. *Nano Lett.* **13**, 1416-1421 (2013).
56. Ross, J. S. et al. Interlayer Exciton Optoelectronics in a 2D Heterostructure p–n Junction. *Nano Lett.* **17**, 638-643 (2017).
57. Shanks, D. N. et al. Interlayer Exciton Diode and Transistor. *Nano Lett.* **22**, 6599-6605 (2022).
58. Ye, T. et al. Room-Temperature Exciton-Based Optoelectronic Switch. *Small* **17**, 2005918 (2021).

59. Wang, K. et al. Electrical control of charged carriers and excitons in atomically thin materials. *Nat. Nanotechnol.* **13**, 128-132 (2018).

60. Perdew, J. P. et al. Atoms, molecules, solids, and surfaces: Applications of the generalized gradient approximation for exchange and correlation. *Phys. Rev. B* **46**, 6671-6687 (1992).

61. Kresse, G. & Furthmüller, J. Efficient iterative schemes for ab initio total-energy calculations using a plane-wave basis set. *Phys. Rev. B* **54**, 11169-11186 (1996).

62. Gonze, X. & Lee, C. Dynamical matrices, Born effective charges, dielectric permittivity tensors, and interatomic force constants from density-functional perturbation theory. *Phys. Rev. B* **55**, 10355-10368 (1997).

63. Togo, A. & Tanaka, I. First principles phonon calculations in materials science. *Scr. Mater.* **108**, 1-5 (2015).

64. Peng, H., Yang, Z.-H., Perdew, J. P. & Sun, J. Versatile van der Waals density functional based on a meta-generalized gradient approximation. *Phys. Rev. X* **6**, 041005 (2016).

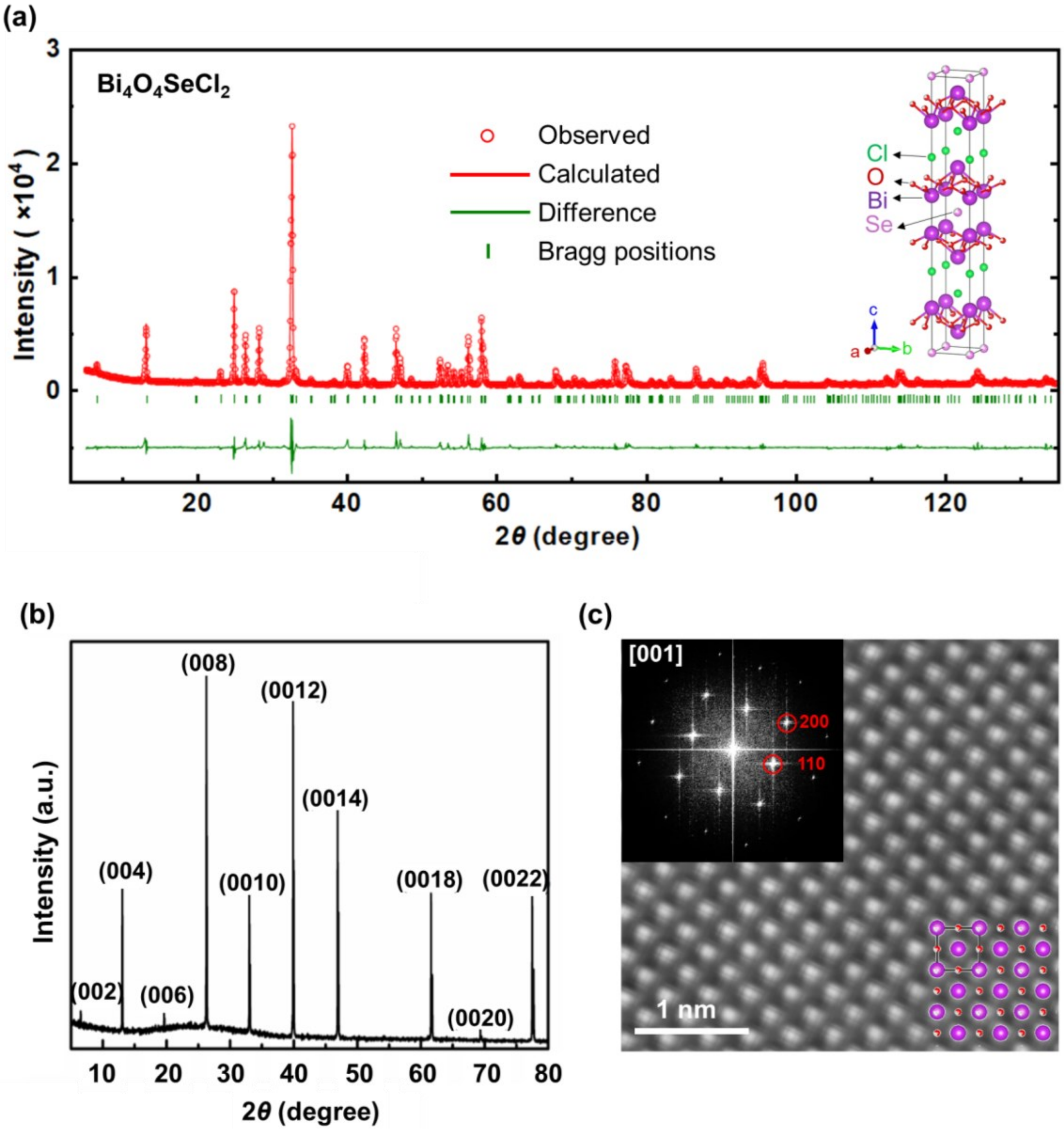

**Fig. 1. Crystal-structure characterization of the $Bi_4O_4SeCl_2$ single crystals. a** Rietveld refinements of powder X-ray diffraction pattern of $Bi_4O_4SeCl_2$. The difference (green line) between the observed (red circles) and the fitted patterns (red line) is shown under the calculated Bragg reflection positions (green vertical bars). The inset shows the crystal structure of $Bi_4O_4SeCl_2$. **b** Single-crystal X-ray diffraction pattern of $Bi_4O_4SeCl_2$. Only (00*l*) diffraction peaks can be observed. **c** The High-angle annular dark-field image of $Bi_4O_4SeCl_2$ along the [001] zone axis. The upper left inset shows the corresponding fast Fourier transformation images. The Bi and O atoms in the unit cell are superimposed on the spots in **c**.

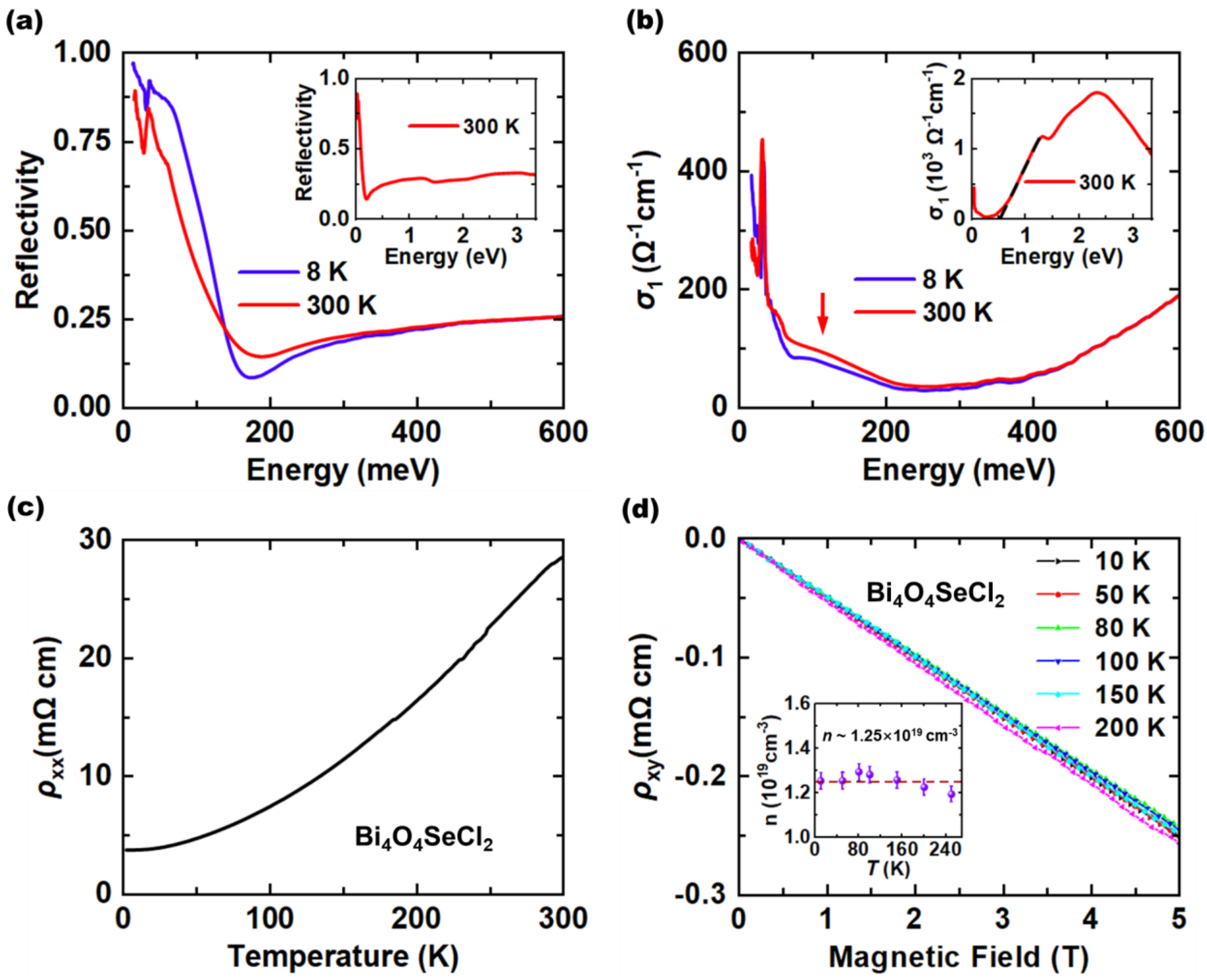

**Fig. 2. Optical spectra and charge-transport data of the $Bi_4O_4SeCl_2$ single crystals. a** Reflectance spectra $R(\omega)$ measured with the electric field $\boldsymbol{E}$ // *ab*-plane of the $Bi_4O_4SeCl_2$ single crystal at different temperatures. The inset shows the $R(\omega)$ measured at 300 K over a broad energy range. **b** Real parts $\sigma_1(\omega)$ of the *ab*-plane optical conductivity at 8 K and 300 K. The red arrow in **b** indicates a peak-like feature around ~ 125 meV in the $\sigma_1(\omega)$. The inset displays the $\sigma_1(\omega)$ at 300 K over a broad frequency range. The black dashed line in the inset of **b** is used for the linear extrapolation of the energy gap. **c** Temperature dependence of the *ab*-plane electrical resistivity $\rho_{xx}$ of the $Bi_4O_4SeCl_2$ single crystal. **d** Magnetic-field dependence of the Hall resistivity $\rho_{xy}$ of the $Bi_4O_4SeCl_2$ single crystal at various temperatures. The inset shows the charge-carrier densities $n$ at different temperatures (the error bars are given by the linear fit).

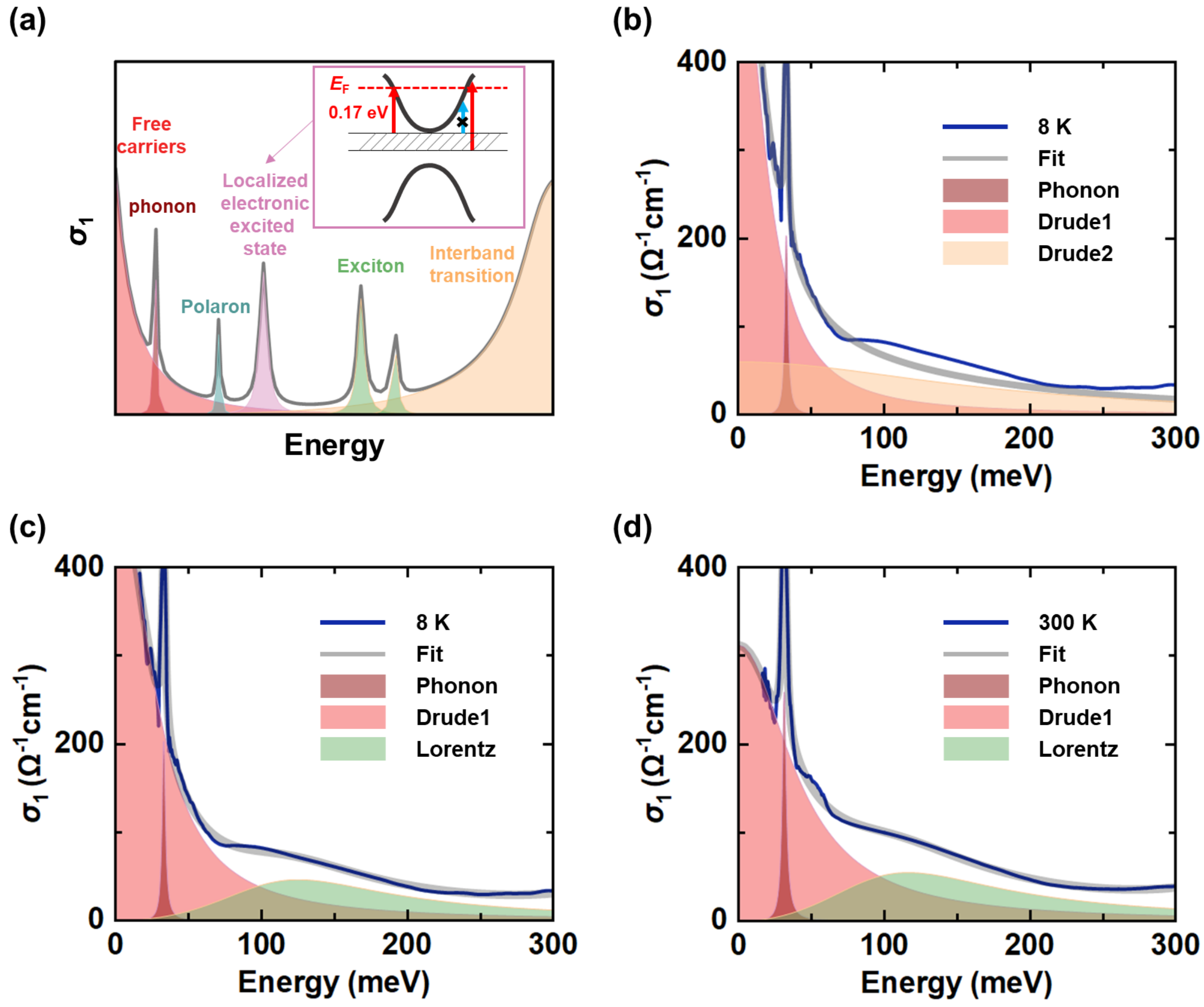

**Fig. 3. Drude–Lorentz fit to the peak-like feature around ~ 125 meV in the optical conductivity spectra of the $Bi_4O_4SeCl_2$ single crystals. a** Schematic drawing of the $\sigma_1(\omega)$ of a material. The inset shows the allowed optical transitions from the localized electronic states to the unoccupied states above the Fermi level. **b** Drude-Lorentz fit to the $\sigma_1(\omega, T = 8\ K)$ of $Bi_4O_4SeCl_2$ using two Drude components. **c** Drude-Lorentz fit to the $\sigma_1(\omega, T = 8\ K)$ of $Bi_4O_4SeCl_2$ using a single Drude component. **d** Drude-Lorentz fit to the $\sigma_1(\omega, T = 300\ K)$ of $Bi_4O_4SeCl_2$ using one Drude component.

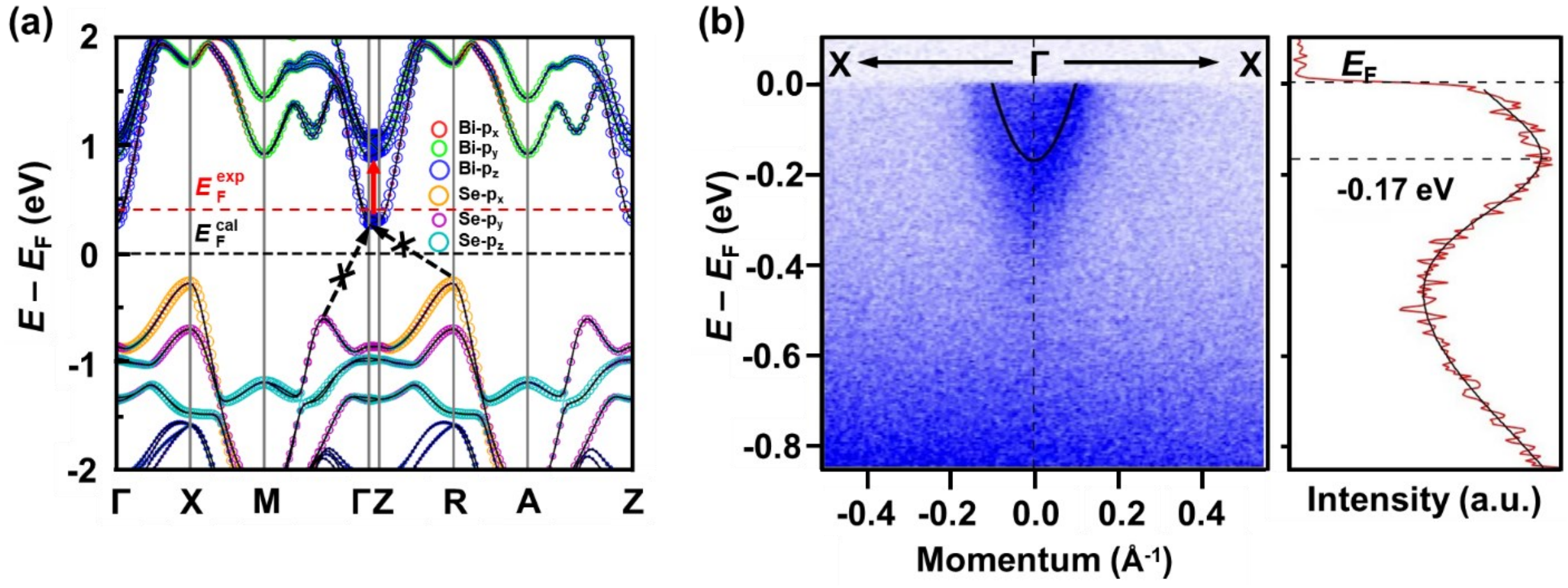

**Fig. 4. Calculated and measured electronic band structures of $Bi_4O_4SeCl_2$. a** Electronic band structure of $Bi_4O_4SeCl_2$ obtained by first-principle calculations. The *black* dashed horizontal line shows the Fermi levels derived from first-principle calculations. The red arrow indicates the allowed direct interband transition at the Γ point. The indirect interband transitions denoted by the black arrows are forbidden. **b** Energy dispersion for the conduction band of $Bi_4O_4SeCl_2$ along the Γ-X direction measured by ARPES (left panel). Energy distribution curve at the Γ point (right panel). The Fermi level above the conduction bottom in **b** is about 0.17 eV. The effective mass of the conduction band in **b** is ~ 0.278 ± 0.01 $m_0$. The *red* dashed horizontal line in a) indicates the Fermi level measured by ARPES.

**Table 1.** The parameters of the Drude-Lorentz fit to the $\sigma_1(\omega, T = 8\text{ K})$ with two Drude components.

| $j$ | $\omega_{pj}$ (meV) | $\Gamma_{Dj}$ (meV) | $\omega_j$ (meV) | $S_j$ (meV) | $\Gamma_j$ (meV) |
|---|---|---|---|---|---|
| 1 | 289 | 20 | - | - | - |
| 2 | 279 | 174 | - | - | - |
| 1 | - | - | 920 | 827 | 402 |
| 2 | - | - | 1227 | 1621 | 561 |
| 3 | - | - | 1707 | 885 | 481 |
| 4 | - | - | 2369 | 4848 | 1842 |

**Table 2.** The parameters of the Drude-Lorentz fit to the $\sigma_1(\omega, T = 8\text{ K})$ with one Drude component.

| $j$ | $\omega_{pj}$ (meV) | $\Gamma_{Dj}$ (meV) | $\omega_j$ (meV) | $S_j$ (meV) | $\Gamma_j$ (meV) |
|---|---|---|---|---|---|
| 1 | 334 | 26 | - | - | - |
| 1 | - | - | 125 | 225 | 146 |
| 2 | - | - | 920 | 827 | 402 |
| 3 | - | - | 1227 | 1621 | 561 |
| 4 | - | - | 1707 | 885 | 481 |
| 5 | - | - | 2369 | 4848 | 1842 |

**Table 3.** The parameters of the Drude-Lorentz fit to the $\sigma_1(\omega, T = 300$ K) with one Drude component.

| $j$ | $\omega_{pj}$ (meV) | $\Gamma_{Dj}$ (meV) | $\omega_j$ (meV) | $S_j$ (meV) | $\Gamma_j$ (meV) |
|---|---|---|---|---|---|
| 1 | 314 | 43 | - | - | - |
| 1 | - | - | 116 | 248 | 150 |
| 2 | - | - | 920 | 827 | 402 |
| 3 | - | - | 1227 | 1621 | 561 |
| 4 | - | - | 1707 | 885 | 481 |
| 5 | - | - | 2369 | 4848 | 1842 |

**Table 4.** The energies, irreducible representations and species of the calculated infrared-active (IR) and Raman-active phonon modes of $Bi_4O_4SeCl_2$.

| Symmetry | Activity | Energy (meV) | Species |
|---|---|---|---|
| $E_g$ | Raman | 3.93 | TO |
| $A_{1g}$ | Raman | 8.71 | LO |
| $E_u$ | IR | 9.59 | LO |
| $E_g$ | Raman | 9.95 | TO |
| $E_u$ | IR | 11.26 | LO |
| $A_{2u}$ | IR | 14.67 | TO |
| $E_u$ | IR | 15.84 | LO |
| $A_{2u}$ | IR | 17.33 | LO |
| $A_{1g}$ | Raman | 20.90 | LO |
| $E_g$ | Raman | 21.07 | LO |
| $A_{2u}$ | IR | 23.47 | LO |
| $A_{1g}$ | Raman | 26.25 | LO |
| $E_u$ | IR | 38.78 | LO |
| $E_g$ | Raman | 39.41 | TO |
| $B_{2u}$ | Raman | 49.75 | LO |
| $B_{1g}$ | Raman | 49.91 | LO |
| $A_{2u}$ | IR | 57.22 | TO |
| $E_u$ | IR | 57.89 | TO |
| $E_g$ | Raman | 58.23 | LO |
| $A_{1g}$ | Raman | 67.02 | LO |